\documentclass[10pt,prd,twocolumn,tightlines,showpacs,preprintnumbers,aps,
floatfix,amsmath,amssymb]{revtex4}
\begin{document}
\input psfig.tex
\title{Stochastic Quantisation and Non-Equilibrium Statistical Mechanics}
\author{Jayanta K. Bhattacharjee$^a$ and Debashis Gangopadhyay$^b$}
\affiliation {$^a$Indian Association For The Cultivation Of Science, Jadavpur, 
Kolkata-700032, India , tpjkb@mahendra.iacs.res.in, $^b$S.N.Bose National Centre
For Basic Sciences,JD-Block, Sector-III,Salt Lake City,Kolkata-700098,India,
debashis@boson.bose.res.in}
\date{\today}
\begin{abstract}
The stochastic quantisation technique of Parisi and Wu is extended to study
non-equilibrium statistical mechanics. We show that this scheme is capable 
of handling white as well as coloured noises.

\end{abstract}
\maketitle
A quarter of a century back , Parisi and Wu [1] proposed a method of quantisation
based on a stochastic  Langevin dynamics of a physical system in a fifth time $\tau$. 
They showed that at the perturbative level, the usual quantum field thery (at the 
level of the correlation functions) was recovered in the $\tau\rightarrow\infty$ 
of this dynamics. In this work, we extend this technique to study non-equilibrium
statistical mechanics. 

To study systems out of equilibrium, the usual practice is  
to use response fields [2].
These are used to determine propagators and correlators.
As these are unrelated, one has to separarately calculate them.

In the stochastic quantisation scheme propagators and correlation functions
in the  fictitous "time" dimension are related in the usual way. 
Moreover, the equilibrium limit , {\it viz.} $\tau\rightarrow\infty$ 
gives back the real time propagators and correlation functions. 
This way of looking at the problem allows us the luxury of a 
fluctuation-dissipation in the fictitous time and what is more
important, our procedure works unaltered even when the real time 
noise source is coloured.

{\bf 1. Parisi-Wu Stochastic Quantisation }

We begin with a brief review. Euclidean quantum field theoretic correlation functions 
for the field $\phi$ corresponding to an action $S[\phi]$ are given by [1,3]
$$<0\vert T\phi(x_{1})\phi(x_{2})...\phi(x_{l})\vert 0>$$
$$={\int D\phi[\phi(x_{1})\phi(x_{2})..\phi(x_{l})]e^{-S[\phi]}\over \int D\phi e^{-S[\phi]}}\eqno(1)$$
Parisi and Wu proposed the following alternative method:

(a)Introduce an extra fictitous "time" dimension $\tau$ in addition to the four spacetime
$x^{\mu}$ and postulate a Langevin dynamics of the field $\phi$ in the "time" $\tau$ as
$${\partial \phi(x,t,\tau)\over\partial \tau} = - {\delta S\over\delta \phi} + \eta (x,t,\tau)\eqno(2)$$
where  $\eta$ is a gaussian random variable satisfying $<\eta(x,t,\tau)>_{\eta}=0$;
$<\eta(x,t,\tau)\eta(x',t',\tau')>_{\eta}= 2\delta (x-x')\delta (t-t')\delta (\tau - \tau ')$

(b)Evaluate the stochastic average of the fields $\phi_{\eta}$ satisfying $(2)$ {\it viz.}
$$<\phi(x_{1},\tau_{1})\phi(x_{2},\tau_{2})...\phi(x_{l},\tau_{l})>_{\eta}$$
(c)Put $\tau_{1}=\tau_{2}=...\tau_{l}$ and take the limit $\tau_{1}\rightarrow\infty$.
Then one has
$$lim_{\tau_{1}\rightarrow\infty}<\phi(x_{1},\tau_{1})\phi(x_{1},\tau_{1})...\phi(x_{l},\tau_{1})>_{\eta}$$
$$={\int D\phi[\phi(x_{1})\phi(x_{2})..\phi(x_{l})]e^{-S[\phi]}\over \int D\phi e^{-S[\phi]}}\eqno(3)$$
Thus the real time correlation functions are obtained in the equilibrium limit of the 
fictitous time $\tau$.Why this is possible is understood from the associated Fokker-Planck dynamics.
The probability of finding the system in the configuration $\phi$ at time $t$ is $P(\phi,t)$ 
and the evolution of $P$ in time $t$ is governed by the Fokker-Planck equation 
$${\partial P \over\partial t}={\delta^{2} P\over\delta \phi^{2}} + 
{\delta\over\delta \phi}\Bigl(P {\delta S\over\delta \phi}\Bigr)\eqno(4)$$ 
The above equation can be recast into Schroedinger-type equation 
${\partial\Psi\over\partial t}=-2H^{FP}\Psi$ 
with $\Psi\equiv P(\phi,t)e^{S[\phi]/2}$ and 
$$H^{FP}= -(1/2){\delta^{2}\over\delta \phi{^2}}+(1/8)\Bigl({\delta S\over\delta \phi}\Bigr)^{2}
-(1/4){\delta^{2} S\over\delta \phi^{2}}\eqno(5)$$
The Fokker-Planck hamiltonian $H^{FP}$ is positive semi-definite with 
ground state energy zero and ground state $\Psi_{0}=e^{-S[\phi]/2}$ and 
$lim_{t\rightarrow\infty} P(\phi,t)=constant\enskip e^{-S[\phi]/2}$.

{\bf 2.The Edwards-Wilkinson (EW) Model}

One of the earliest models for surface growth is the linear  Edwards-Wilkinson
model [4] described by the equation of motion 
$${\partial h(x,t)\over\partial t}= \nu\nabla ^ {2} h(x,t)+\beta(x,t) + c\eqno(6a) $$
where $h$ is the surface's position,$\beta$ the stochastic white  noise and $c$ the
constant average velocity of the propagating surface.$\beta$ satisfies
 $$<\beta (x_{1},t_{1})\beta (x_{2},t_{2})>
= 2\epsilon\delta (x_{2}-x_{1})\delta (t_{2}-t_{1})\eqno(6b)$$
It can be shown that corresponding to the Langevin equation 
$${\partial h(x,t)\over\partial t}= -{\delta S\over\delta h} + \beta (x,t)\eqno(7)$$
the Hamiltonian 
$H_{EW}=\nu/2 \int dx (\nabla h)^{2}$ gives back equation $(6a)$.  
We apply the technique first to the Edwards-Wilkinson model which is a linear 
theory.The model is described in real time $t$ by the equations $(7)$ or $(6)$.

The probability distribution associated with $\beta$ is 
$$P(\beta)\propto e^{-{1\over 2\epsilon}\int d^{D}xdt \beta(x,t)^{2}}
= e^{-{1\over 2\epsilon}\int d^{D}xdt 
({\partial h\over\partial t}-\nu\nabla^{2} h)^{2}}\eqno(8a)$$
$$=e^{-{1\over 2\epsilon}\int 
{d^{D}k \over (2\pi)^{D}}{d\omega\over 2\pi}
(\omega^{2}+\nu^{2}k^{4})h(k,\omega)h(-k,-\omega)}=e^{-S}$$
where
$$S= {1\over 2\epsilon}\int {d^{D}k \over (2\pi)^{D}}{d\omega\over 2\pi}
 (\omega^{2}+\nu^{2}k^{4})h(k,\omega)h(-k,-\omega)\eqno(8b)$$
Introduce the fictitous time "$\tau$" such that 
$${\partial h(k,\omega,\tau)\over\partial \tau} 
= - {\delta S\over\delta h } 
+ \eta (k,\omega,\tau)\eqno(9a)$$
$$<\eta(k,\omega,\tau)\eta(k',\omega',\tau')
=2\delta(\tau-\tau')\delta(k+k')\delta(\omega+\omega')\eqno(9b)$$
$(9b)$ ensures that the "fluctuation-dissipation" theorem is satisfied 
in $\tau$(even if it may not hold in real time $t$) and that as 
$\tau\rightarrow\infty$,
$P(k,\omega,\tau)\rightarrow P(\beta)$ ,the equilibrium distribution.
To calculate the propagator, there are two approaches:

{\bf (a) From $\tau$-variable Langevin Equation}

The equation is 
$${\partial h(x,t,\tau)\over\partial \tau} = \nu \nabla^{2} h(x,t,\tau) + \eta (x,t,\tau)\eqno(10)$$
The solution in momentum space is 
$$h(k,\tau) = \int_{0}^{\infty} d\tau' e^{-\nu k^{2}(\tau-\tau')}\eta (k,\tau')$$
and the propagator is 
$$<h(k,\omega,\tau)h(-k,-\omega,\tau')$$
$$={1\over\nu k^{2}}
[e^{\nu k^2\vert \tau-\tau'\vert}-e^{-\nu k^2\vert \tau+\tau'\vert}]
\delta (\tau-\tau')\eqno(11)$$
In the limit of $\tau=\tau'\rightarrow\infty$ all the usual EW results are obtained.

{\bf (b) From the Fokker-Planck action}

In the functional integral approach to stochastic quantisation [3], one has a 
"euclidean" lagrangian $L^{FP}$ corresponding to $H^{FP}$
$$L^{FP}=(1/2)\Biggl({\partial h\over\partial \tau}\Biggr)^2
+ (1/8)\Biggl({\delta S\over\delta h}\Biggr)^2
- (1/4)\Biggl({\delta^2 S\over\delta h^2}\Biggr)\eqno(12)$$
and the Fokker-Planck action is 
$$S^{FP}=\int dx\enskip dt\enskip d\tau L^{FP}\eqno(13a)$$
In $(k,\omega,\omega_{\tau})$ space, 
this action takes the form
$$S^{FP\enskip '}=\int ({dk\over 2\pi})({d\omega\over 2\pi})({d\omega_{\tau}\over 2\pi})$$
$$h(k,\omega,\omega_{\tau})
\Biggl(\omega_{\tau}^2 +  + [(\omega^2+\nu ^{2}k^{4})]^{2}\Biggr) 
h(-k,-\omega,\omega_{\tau})\eqno(13b)$$
Thus the height-height correlation function in configuration space is 
$$<h(x,t,\tau)h(0,t',\tau ')>=\int_{-\infty}^{\infty}({dk\over 2\pi})({d\omega\over 2\pi})({d\omega_{\tau}\over 2\pi})$$
$${e^{ikx - i\omega (t-t') - i\omega_{\tau} (\tau - \tau ')}\over
\omega_{\tau}^2 + (\omega^2+\nu ^{2}k^{4})^{2}}\eqno(14)$$ 
In our treatment the Langevin dynamics is effectively taken to start at
$\tau=-\infty$ so that at any finite time $\tau$, which is infinitely 
removed from the starting time, we are already in the equilibrium distribution.
Putting $t=t', \tau=\tau'$ and completing the integrations give
$$<h(0)h(x)>\sim x^{1}\eqno(15)$$
so that the index $\alpha=1/2$.

We now consider the scaling behaviour for the height variable $h$.
Assume ,for some length scale $L$, that
$t\sim L^z , h\sim L^\alpha$ in $D$ space dimensions.Note that we 
regard $\tau$ as a parameter only and hence its scaling properties are meaningless.
This is more so justified as $\tau\rightarrow\infty$ gives us the usual 
real time correlation functions. It readily follows that $\alpha=1 - D/2, z=2$ 
for $L^{FP\enskip '}$ as in equation $(17)$. So for $D=1$ , $\alpha= 1/2$.

Note that if we had started from a modified Langevin equation of the form.
$$\Biggl({\partial h\over\partial t}\Biggr)= P\Biggl({\delta S\over\delta h}\Biggr) +\eta$$
the only change would have been a factor $P$ in front of the term proportional to
$\Biggl({\delta^{2} S\over\delta h^{2}}\Biggr)$ in the action.
But this term does not contribute to the model considered here and so there will 
not be any  change in our conclusions.

{\bf 3. Edwards-Wilkinson model with with coloured Gaussian noise}

For coloured Gaussian noise (in real time) the EW model is described by the equations
$${\partial h(x,t)\over\partial t}= \nu\nabla ^ {2} h(x,t)+\beta(x,t) \eqno(16) $$
$$<\beta (k_{1},\omega_{1})\beta (k_{2},\omega_{2})>
={2\epsilon\over k_{1}^{2\rho}}\delta (k_{1}+k_{2})\delta(\omega_{1}+\omega_{2})\eqno(17)$$
$$P(\beta)\propto 
=e^{-{1\over 2\epsilon}
\int {d^D k\over (2\pi)^D} {d\omega\over 2\pi}
k^{2\rho}(\omega^{2}+\nu^{2}k^{4})h(k,\omega)h(-k,-\omega)}\eqno(18)$$
Writing the fictitous time Langevin equation as
$${\partial h(k,\omega,\tau)\over\partial\tau}
=-k^{2\rho}(\omega^{2}+\nu^{2}k^{4})h(k,\omega,\tau) + \eta(k,\omega,\tau)\eqno(19)$$
where $\eta$ satisfies $(9b)$.The Fokker-Planck equation corresponding to $(9a)$ is
$${\partial P\over\partial \tau}
={\partial\over\partial h(k,\omega)}(P {\partial S\over\partial h(-k,-\omega)})
+{\partial^{2} P\over\partial h(-k,w)\partial h(-k,\omega)}$$
The Fokker-Planck hamiltonian is now
$$H^{FP}= ({\delta h\over\delta \tau})^{2}+({1\over 2})\Bigl({\delta S\over\delta h}\Bigr)^{2}
-({1\over 4}){\delta^{2} S\over\delta h^{2}}\eqno(20)$$
So by going over to the fictitous time , it is always possible to write
a Fokker-Planck hamiltonian even if the original equation in real time 
does not have the structure ${\partial h\over\partial t}=-{\delta\tilde S\over \delta h} +\eta$

Proceeding as before the propagator in coordinate space in the limit 
$\tau_{1}=\tau_{2}\rightarrow\infty$ is 
$$<h(0)h(x)>=\int {dk e^{ikx}\over 2\pi}\int {dw\over 2\pi}{dw_{\tau}\over 2\pi}$$
$${1\over \omega_{\tau}^{2} + [k^{2\rho}(\omega^{2}+\nu^{2}k^{4})]^{2}}\sim x^{1+2\rho}$$
Therefore the index $\alpha=1/2 + \rho$.
\begin{figure}
\vbox{
\vskip 1.0cm
\hskip 0.5cm
\centerline{
\psfig{figure=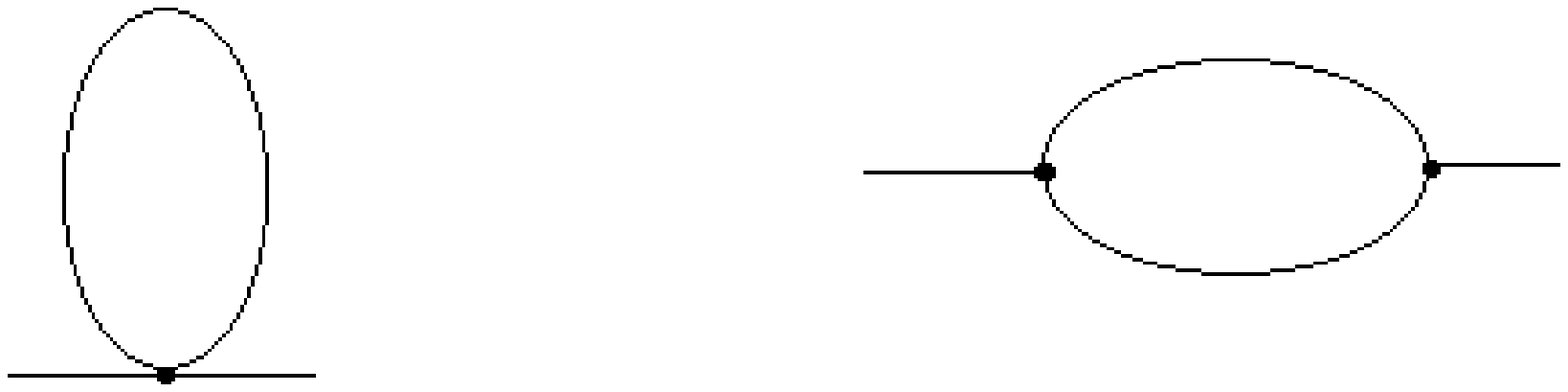,height=6truecm,width=12truecm}}}
\vspace{0.0cm}
\caption{
}
\end{figure}
Now consider the model of Medina et al[ref.6]
$${\partial h(x,t,\tau)\over\partial t} = \nu \nabla^{2} h(x,t,\tau) 
-(\lambda/2) (\nabla h)^{2}+ \beta (x,t)\eqno(21)$$
with $\beta$ satisfying $(17)$.The action is 
$$S=(1/2)\int {d^{D}k\over (2\pi)^{D}}{d\omega\over 2\pi}
\Biggl( k^{2\rho}(\omega ^2+\nu ^2 k^4)h(k,\omega)h(-k,-\omega)$$
$$+\lambda k^{2\rho}\int{d^{D}p\over (2\pi)^{D}}{d\omega '\over 2\pi}$$
$$(i\omega+\nu k^2){\bf p.(k-p)}h({\bf p},\omega ')h({\bf k-p},\omega-\omega ')h({\bf -k,-\omega})$$
$$-\lambda k^{2\rho}\int{d^{D}p\over (2\pi)^{D}}{d\omega '\over 2\pi}$$
$$(-i\omega+\nu k^2){\bf p.(k+p)}h({\bf p},\omega ')h({\bf -k-p},-\omega-\omega ')h({\bf k,\omega})$$
$$-\lambda k^{2\rho}\int {d^{D}p\over (2\pi)^{D}}{d^{D}q\over (2\pi)^{D}}{d\omega_{1}\over 2\pi}
{d\omega_{2}\over 2\pi}$$
$$[{\bf p.(k-p)}][{\bf q.(k-q)}]h({\bf p},\omega_{1})
h({\bf k-p},\omega-\omega_{1})$$
$$h({\bf q},\omega_{2})h({\bf k-q},\omega-\omega_{2})\Biggr)\eqno(22)$$

In fictitous time the Langevin eqution is 
$${\partial h(k,\omega,\tau)\over\partial \tau} 
= - {\delta S\over\delta h(-k,-\omega,\tau)} 
+ \eta (k,\omega,\tau)\eqno(23a)$$
$$<\eta(k,\omega,\tau)\eta(k',\omega',\tau')
=2\delta(\tau-\tau')\delta(k+k')\delta(\omega+\omega')\eqno(23b)$$
$(23b)$ ensures that the "fluctuation-dissipation" theorem is satisfied 
in $\tau$.The evolution to the equilibrium limit is governed by 
$S^{FP\enskip '}$ constructed from the action $S$ as in $(22)$
and using $(12)$ and $(13)$.
The fluctuation-dissipation theorem relates the connected 
two point function to the imaginary part of the response function i.e.
$$C = {1\over\omega_{\tau}}Im G\eqno(24)$$
where $C$ refers to connected part of the two-point function in fictitous time $\tau$
and $G$ refers to the full response function.
The full response function has the structure 
$$G= {1\over -i\omega_{\tau}+k^{2\rho}(\omega^{2}+k^{2z})}$$
while 
$$C={1\over k^{4\rho}(\omega^{2}+k^{2z})^{2})+\omega_{\tau}^{2}}=G^{0}G^{0 *}$$

Upto one loop level the relevant diagrams are in fig.1. Of these the first
one is like a mass insertion vertex (of perturbation order $O(\lambda)$)
and hence ignorable.However, the second one ($O(\lambda^{2})$)
depends on the external momenta.Now considering upto one loop level
$$G^{-1}\approx G^{-1 (0)}-\Sigma$$
where $\Sigma$ contains the relevant one loop contribution, i.e. the second
diagram. Denoting by $V$ the contribution from the vertex
$$\Sigma=\int {d^{D}p\over (2\pi)^{D}}
{d\omega\over (2\pi)} {d\omega_{\tau}\over (2\pi)}VVGC$$

Now consider the momentum behaviour.
Noting that $\nu k^{2} \simeq k^{z}$,
the second diagram gives $\Sigma$ behaviour
as $\sim G^{-1}\sim k^{2\rho+2z}$.The vertex $V$ behaves as 
$\sim k^{2\rho}\nu k^{2}.k^{2}\sim k^{2\rho+z+2}$,
while dimensional matching implies that $\omega\sim k^{z}$.
Therefore $k^{2\rho+2z}=k^{D+z+4\rho+2z+4-4\rho-4z}$.
This gives $z={1\over 3}(D+4-2\rho)$.

This result agrees exactly with what is found in 
refs. (4) and (5). Since we expect results to be more singular
with $\rho > 0 $, we expect the validity of our calculation 
only for that range of $\rho $ for which $ z $ is smaller than the 
corresponding $z$ for the KPZ problem. For $ D=1$, this restricts
the validity of our analysis to $\rho >1/4$, which is exactly
what was found in ref.(5). In summary, then we have a technique 
where the out-of-equilibrium system with a coloured noise
can be handled by techniques of equilibrium statistical physics.

\end{document}